\documentclass[aps,prd,onecolumn,preprintnumbers,groupedaddress,showpacs,nofootinbib,amssymb]{revtex4-2}
\usepackage{graphicx,color}
\usepackage{amsmath}
\usepackage{amssymb}
\usepackage{amsfonts}
\usepackage{bm}
\usepackage{cancel}
\usepackage{slashed}

\allowdisplaybreaks[4]

\begin{document}

\tolerance=5000
\title{Strong Einstein-Hilbert Gravity Inflation and ACT Phenomenology}
\author{V.K. Oikonomou,$^{1}$}\email{voikonomou@gapps.auth.gr;v.k.oikonomou1979@gmail.com}
\author{Eleni I. Manouri,$^{1}$}\email{elenimanouri21@gmail.com;emanouri@physics.auth.gr}
\author{Georgios Konstantellos,$^{1}$}\email{konstantellosgiorgos9@gmail.com}
\affiliation{$^{1)}$ Department of Physics, Aristotle University
of Thessaloniki, Thessaloniki 54124, Greece}

\tolerance=5000

\begin{abstract}
In this work we study rescaled effective single scalar field
theories, and we confront these with the ACT constraint on the
spectral index of the scalar primordial perturbations and the
updated BICEP/Planck constraint on the tensor-to-scalar ratio.
Rescaled scalar theories of gravity may be the result of an
effective $f(R,\phi)$ gravity at strong curvature regimes, which
may result on a rescaling of the Einstein-Hilbert term of the form
$\sim \alpha R$. It turns out that canonical scalar field theories
with stronger gravity compared to standard Einstein-Hilbert
gravity can be compatible with the ACT and updated Planck/BICEP
constraints, with stronger gravity meaning that the rescaling
parameter $\alpha$ takes values smaller than unity.
\end{abstract}

\maketitle

\section{Introduction}

During the next ten years the inflationary era will be in the
focus of scientific research via the upcoming observations.
Specifically the ground based stage 4 Cosmic Microwave Background
(CMB) radiation experiments like the Simons observatory
\cite{SimonsObservatory:2019qwx}, and the LiteBird experiment
\cite{LiteBIRD:2022cnt} will directly probe the $B$-mode of the
CMB, which will be a smoking gun signal for the inflationary era.
On the other hand, the inflationary regime will indirectly be
probed by the future gravitational wave experiments
\cite{Hild:2010id,Baker:2019nia,Smith:2019wny,Crowder:2005nr,Smith:2016jqs,Seto:2001qf,Kawamura:2020pcg,Bull:2018lat,LISACosmologyWorkingGroup:2022jok}.
These experiments will seek for stochastic gravitational wave
backgrounds of cosmological origin, thus with small anisotropies.
Such a background was detected in 2023 by NANOGrav and other
Pulsar-Timing-Array experiments
\cite{nanograv,Antoniadis:2023ott,Reardon:2023gzh,Xu:2023wog}, but
it is highly unlikely that an inflationary era by itself can
generate the NANOGrav 2023 stochastic gravitational wave signal
\cite{Vagnozzi:2023lwo,Oikonomou:2023qfz}.

Inflation  \cite{inflation1,inflation2,inflation3,inflation4} is a
speculative theory for the primordial Universe that needs to be
verified experimentally. There are various theories that can
produce an inflationary regime, such as canonical scalar field
theories, which is a theory within Einstein-Hilbert gravity, and
even modified gravity \cite{reviews1,reviews2,reviews3,reviews4},
with the prominent modified gravity theory being $F(R)$ gravity
\cite{Nojiri:2003ft,Capozziello:2005ku,Capozziello:2004vh,Capozziello:2018ddp,Hwang:2001pu,Cognola:2005de,Nojiri:2006gh,Song:2006ej,Capozziello:2008qc,Bean:2006up,Capozziello:2012ie,Faulkner:2006ub,Olmo:2006eh,Sawicki:2007tf,Faraoni:2007yn,Carloni:2007yv,
Nojiri:2007as,Capozziello:2007ms,Deruelle:2007pt,Appleby:2008tv,Dunsby:2010wg,Odintsov:2020nwm,Odintsov:2019mlf,Odintsov:2019evb,Oikonomou:2020oex,Oikonomou:2020qah,Huang:2013hsb,Berry:2011pb,Bonanno:2010bt,Gannouji:2008wt,Oyaizu:2008sr,Oyaizu:2008tb,Brax:2008hh,Cognola:2007zu,Boehmer:2007glt,Boehmer:2007kx,
deSouza:2007zpn,Song:2007da,Brookfield:2006mq,delaCruz-Dombriz:2006kob,Achitouv:2015yha,Kopp:2013lea,Sebastiani:2013eqa,Odintsov:2017hbk,Myrzakulov:2015qaa,Feng:2022vcx},
based on a simple Occam's razor approach. The line of thinking is
simple, Einstein-Hilbert gravity contains the Ricci scalar in the
gravitational Lagrangian density, therefore one may expect that
any modification of Einstein-Hilbert gravity will manifest itself
in the inflationary Lagrangian, containing a function of the Ricci
scalar. Now modified gravity is nowadays more important than never
in the previous years and is expected to affect probably and
mainly the primordial and the late-time eras of our Universe. Let
us ponder on why modified gravity is highly motivated nowadays.
The latest DESI data \cite{DESI:2024uvr} indicated that dark
energy is there for sure and controls the late-time dynamics, but
surprisingly, dark energy does not manifest itself as a
cosmological constant, which is described by the
$\Lambda$-Cold-Dark-Matter model ($\Lambda$CDM), but it is
dynamical and more surprisingly it evolves from a phantom regime
to a quintessential regime
\cite{Lee:2025pzo,Ozulker:2025ehg,Kessler:2025kju,Nojiri:2025low,Vagnozzi:2019ezj}.
This feature definitely cannot be harbored in an Einstein-Hilbert
framework without resorting to tachyon scalars. On the contrary
phantom evolution and phantom transitions can be naturally
produced by modified gravity without resorting to phantom scalars,
see the recent \cite{Nojiri:2025uew,Nojiri:2025low}. Apart from
this striking evidence which does not favor Einstein-Hilbert
motivated models, there always exists the Hubble tension which
challenges the $\Lambda$CDM model
\cite{Pedrotti:2024kpn,Jiang:2024xnu,Vagnozzi:2023nrq,Adil:2023exv,Bernui:2023byc,Gariazzo:2021qtg}.
Therefore, although general relativity seems to be the description
of the Universe at a local astrophysical level and seems to
describe accurately the physics of compact objects, at large
scales it fails to describe the Universe in a consistent way. Thus
modified gravity seems to be a consistent description of the
Universe at large scales. Now, from a quantum physics perspective,
modified gravity is also there. The inflationary era is expected
to be very sensitive to the physics of deep ultra-violet, since
the inflationary era is a classical theory, very close
energetically to the quantum Planck regime. And modified gravity
manifests itself in the scalar field action, since the first
quantum corrections of the scalar field action, when it is
considered in its vacuum configuration namely,
\begin{equation}\label{generalscalarfieldaction}
\mathcal{S}_{\phi}=\int
\mathrm{d}^4x\sqrt{-g}\left(\frac{1}{2}Z(\phi)g^{\mu
\nu}\partial_{\mu}\phi
\partial_{\nu}\phi-\mathcal{V}(\phi)+h(\phi)\mathcal{R}
\right)\, .
\end{equation}
are the following
\cite{Codello:2015mba,Oikonomou:2025htz,Oikonomou:2025ccs,Oikonomou:2020oex},
\begin{align}\label{quantumaction}
&\mathcal{S}_{eff}=\int
\mathrm{d}^4x\sqrt{-g}\Big{(}\Lambda_1+\Lambda_2
\mathcal{R}+\Lambda_3\mathcal{R}^2+\Lambda_4 \mathcal{R}_{\mu
\nu}\mathcal{R}^{\mu \nu}+\Lambda_5 \mathcal{R}_{\mu \nu \alpha
\theta}\mathcal{R}^{\mu \nu \alpha \theta}+\Lambda_6 \square
\mathcal{R}\\ \notag &
+\Lambda_7\mathcal{R}\square\mathcal{R}+\Lambda_8 \mathcal{R}_{\mu
\nu}\square \mathcal{R}^{\mu
\nu}+\Lambda_9\mathcal{R}^3+\mathcal{O}(\partial^8)+...\Big{)}\, ,
\end{align}
where the parameters $\Lambda_i$, $i=1,2,...,6$ are appropriate
dimensionful constants. Thus apparently, modified gravity is also
motivated from a quantum perspective, when the inflationary
Lagrangian is considered.

In this work we shall consider the minimal quantum effects of
terms of the form (\ref{quantumaction}) in the single canonical
scalar field action of Eq. (\ref{generalscalarfieldaction}), so we
shall choose  $Z(\phi)=-1$ and $h(\phi)=1$ in Eq.
(\ref{generalscalarfieldaction}). We shall focus on the case that
the overall effect of modified gravity results to a strong gravity
of the form $\sim \alpha R$, where $\alpha$ is a dimensionless
parameter smaller than unity. Thus during the inflationary era,
Newton's gravitational constant is larger compared to the
post-inflationary eras. This idea of having a rescaled
Einstein-Hilbert framework has been introduced and studied earlier
in the literature, see for example
\cite{Oikonomou:2025htz,Oikonomou:2025ccs,Oikonomou:2020oex}. We
shall study the inflationary phenomenology of various single
scalar field models of interest and we shall compare the results
with the stunning ACT data, that constrain the spectral index of
the scalar perturbations to be, \cite{ACT:2025fju,ACT:2025tim},
\begin{equation}\label{ACT}
n_{s}=0.9743 \pm 0.0034\, .
\end{equation}
We will also compare the inflationary phenomenology with the
updated Planck/BICEP constraints on the tensor-to-scalar ratio
which constrain it to be \cite{BICEP:2021xfz},
\begin{equation}\label{planck}
r<0.036
\end{equation}
at $95\%$ confidence. The striking ACT data has motivated a large
stream of articles aiming to explain how such a spectral index may
be generated, see for example
\cite{Kallosh:2025rni,Gao:2025onc,Liu:2025qca,Yogesh:2025wak,Yi:2025dms,Peng:2025bws,Yin:2025rrs,Byrnes:2025kit,Wolf:2025ecy,Aoki:2025wld,Gao:2025viy,Zahoor:2025nuq,Ferreira:2025lrd,Mohammadi:2025gbu,Choudhury:2025vso,Odintsov:2025wai,Q:2025ycf,Zhu:2025twm,Kouniatalis:2025orn,Hai:2025wvs,Dioguardi:2025vci,Yuennan:2025kde,Kuralkar:2025zxr,Kuralkar:2025hoz,Modak:2025bjv,Oikonomou:2025xms,Oikonomou:2025htz,Odintsov:2025jky,Aoki:2025ywt,Ahghari:2025hfy,McDonough:2025lzo,Chakraborty:2025wqn,NooriGashti:2025gug,Yuennan:2025mlg,Deb:2025gtk,Afshar:2025ndm,Ellis:2025zrf,Yuennan:2025tyx}.
In the following sections, we shall analyze the inflationary
phenomenology of several rescaled single scalar field models and
we shall compare these with the ACT and updated BICEP/Planck
constraints.

Before starting our analysis, let us note that for the background
metric we  choose a flat Friedmann Robertson Walker (FRW) metric,
with line element,
\begin{equation}
    \centering\label{frw}
    d s^2 = - d t^2 + a(t) \sum_{i = 1}^3 d x_i^2,
\end{equation}
where $a(t)$ denotes the scale factor. For the flat FRW metric,
the Ricci scalar is given by,
\begin{equation}
    R = 6 \dot{H} + 12 H^2,
\end{equation}
where $H = \frac{\dot{a}}{a}$ stands for the Hubble parameter.
Moreover, $\kappa=\frac{1}{M_p}=8\pi G$, where $M_p$ is the
reduced Planck mass, and $G$ is Newton's gravitational constant.
Also we shall use natural units, in which $c = \hbar = 1$.

\section{The $f(R)$ Gravity Scalar Field Model and Inflationary Effective Theory}

The gravitational action we shall consider in this article has the
following form,
\begin{equation}
\label{action}\mathcal{S}=\int d^4x\sqrt{-g}\left(\frac{\alpha
R}{2\kappa^2}-\frac{1}{2}g^{\mu \nu}\partial_{\mu}\phi
\partial_{\nu}\phi-V(\phi)\right)\, ,
\end{equation}
where $\kappa^2=8\pi G=\frac{1}{M_p^2}$ and $M_p$ stands for the
reduced Planck mass. Such an action, may effectively originate
from an $f(R)$ gravity scalar field model action of the following
form,
\begin{equation}
\label{action} \centering
\mathcal{S}=\int{d^4x\sqrt{-g}\left(\frac{f(R)}{2\kappa^2}-\frac{1}{2}\partial_\mu\phi\partial^\mu\phi-V(\phi)\right)}\,
,
\end{equation}
with the $f(R)$ gravity having the form,
\begin{equation}\label{frini}
f(R)=R-\gamma  \lambda  \Lambda -\lambda  R \exp
\left(-\frac{\gamma  \Lambda }{R}\right)-\frac{\Lambda
\left(\frac{R}{m_s^2}\right)^{\delta }}{\zeta }\, .
\end{equation}
As it was shown in Ref.
\cite{Oikonomou:2025htz,Oikonomou:2025ccs,Oikonomou:2020oex}, the
$f(R)$ gravity of Eq. (\ref{frini}) at leading order, in the large
curvature regime, takes the following form,
\begin{equation}\label{expapprox}
\lambda  R \exp \left(-\frac{\gamma  \Lambda }{R}\right)\simeq
-\gamma \lambda  \Lambda -\frac{\gamma ^3 \lambda \Lambda^3}{6
R^2}+\frac{\gamma ^2 \lambda  \Lambda ^2}{2 R}+\lambda  R\, ,
\end{equation}
hence, the effective inflationary action during the inflationary
regime is approximately,
\begin{equation}\label{effectiveaction}
\mathcal{S}=\int
d^4x\sqrt{-g}\left(\frac{1}{2\kappa^2}\left(\alpha R+ \frac{\gamma
^3 \lambda \Lambda ^3}{6 R^2}-\frac{\gamma ^2 \lambda \Lambda
^2}{2 R}-\frac{\Lambda}{\zeta
}\left(\frac{R}{m_s^2}\right)^{\delta
}+\mathcal{O}(1/R^3)+...\right)-\frac{1}{2}\partial_\mu\phi\partial^\mu\phi-V(\phi)\right)\,
,
\end{equation}
where $\alpha=1-\lambda$. The gravitational action of Eq.
(\ref{effectiveaction}) is a rescaled version of Einstein-Hilbert
canonical scalar field theory. Hence, the starting point of our
analysis is the following rescaled canonical scalar field action,
\begin{equation}
    \centering\label{act}
    S = \int d^4 x \sqrt{- g}\left(\frac{\alpha R}{2 \kappa^2} - \frac{1}{2}g^{\mu \nu} \partial_\mu \phi \partial_\nu \phi - V(\phi) \right),
\end{equation}
with$\alpha$ is a dimensionless parameter which it will prove to
take values in the range $0 < \alpha < 1$ in order to ensure
compatibility with the ACT data. The scalar field equation of
motion is the following,
\begin{equation}
    \centering\label{eomphi}
    \ddot{\phi} + 3 H \dot{\phi} + V' = 0\, .
\end{equation}
Also by varying the original action (\ref{act}) with respect to
the metric tensor, we obtain,
\begin{equation}
    \centering\label{eom}
    \frac{\alpha}{\kappa^2}\left(R_{\mu \nu} - \frac{1}{2}R g_{\mu \nu}\right) = \partial_\mu \phi \partial_\nu \phi - g_{\mu \nu}\left(\frac{1}{2}g^{\rho \sigma}\partial_\rho \phi \partial_\sigma \phi +V(\phi) \right).
\end{equation}
For the FRW metric, the field equations (\ref{eom}) yield the
following Friedmann and Raychaudhuri equations,
\begin{equation}
    \centering\label{fr1}
    \frac{3\alpha}{\kappa^2} H^2 =\frac{1}{2}\dot{\phi}^2 + V(\phi),
\end{equation}
\begin{equation}
    \centering\label{fr2}
    \frac{2 \alpha}{\kappa^2}\dot{H} =-\dot{\phi}^2.
\end{equation}
The slow-roll condition for the inflationary era is imposed,
\begin{equation}
    \centering\label{inf}
    V(\phi) \gg \dot{\phi}^2,
\end{equation}
and by combining Eqs. (\ref{fr1}) and (\ref{fr2}),  we have,
\begin{equation}
    \centering\label{cond}
    \frac{|\dot{H}|}{H^2} \ll 1.
\end{equation}
The first slow-roll parameter is defined as follows,
\begin{equation}
    \centering\label{eps1}
     \epsilon_1 = -\frac{\dot{H}}{H^2},
\end{equation}
and thus Eq. (\ref{fr1}), namely the Friedmann equation, becomes,
\begin{equation}
    \centering\label{h2}
    H^2 = \frac{\kappa^2}{3 \alpha}V(\phi).
\end{equation}
The condition which ensures that the duration of the inflationary
era is sufficiently long, has the following form,
\begin{equation}
    \centering\label{time}
    |\ddot{\phi}| \ll 3 H |\dot{\phi}|,
\end{equation}
and also the second slow-roll parameter is defined as follows,
\begin{equation}
    \centering\label{eps2}
     \epsilon_2 = \frac{\ddot{\phi}}{H \dot{\phi}}.
\end{equation}
Due to Eq. (\ref{time}), from Eq. (\ref{eomphi}) we get,
\begin{equation}
    \centering\label{dphi}
    \dot{\phi} = - \frac{V'}{3 H}\, ,
\end{equation}
and also,
\begin{equation}
    \centering\label{ddphi}
    \ddot{\phi} = - \frac{\dot{H}}{H}\dot{\phi} -V'' \frac{\dot{\phi}}{3 H}.
\end{equation}
By combining Eqs. (\ref{fr2}), (\ref{h2}), (\ref{dphi}),
(\ref{ddphi}), (\ref{eps1}) and (\ref{eps2}), the slow-roll
parameters for the single canonical scalar field theory become,
\begin{equation}
    \centering\label{eps1 to eps}
     \epsilon_1 = \alpha  \epsilon\, ,
\end{equation}
and
\begin{equation}
    \centering\label{eps2 to eta}
     \epsilon_2 = -\alpha \eta +  \epsilon_1,
\end{equation}
with,
\begin{equation}
    \centering\label{eps}
     \epsilon = \frac{1}{2 \kappa^2}\frac{V'^2}{V^2},
\end{equation}
\begin{equation}
    \centering\label{eta}
    \eta = \frac{1}{\kappa^2}\frac{V''}{V} ,
\end{equation}
Also, the $e$-foldings number is defined as follows,
\begin{equation}
    \centering\label{efold}
    N(\phi) = \int_t^{t_{end}} H d t,
\end{equation}
with $t_{end}$ denoting the end of the inflationary era, and by
combining (\ref{h2}) and (\ref{dphi}), Eq. (\ref{efold}) becomes,
\begin{equation}
    \centering\label{N}
    N(\phi) = \frac{\kappa^2}{\alpha} \int_{\phi_{end}}^\phi \frac{V}{V'}d \phi,
\end{equation}
where $\phi_{end}$ is the scalar field value at the end of the
inflationary regime. The spectral index of the scalar
perturbations is,
\begin{equation}
    \centering\label{spec}
    n_s - 1 = -4\epsilon_1 -2\epsilon_2 ,
\end{equation}
thus by substituting (\ref{eps1 to eps}) and (\ref{eps2 to eta}),
it takes the form,
\begin{equation}
    \centering\label{ns}
    n_s = 1 + 2 \alpha \eta - 6 \alpha  \epsilon.
\end{equation}
In addition, the tensor-to-scalar ratio takes the form,
\begin{equation}
    \centering\label{ttsr}
    r = 8 \kappa^2 \frac{\dot{\phi}^2}{H^2},
\end{equation}
which in our case reduces to,
\begin{equation}
    \centering\label{r}
    r = 16 \alpha  \epsilon.
\end{equation}
In the next section we shall use the formalism developed in this
section and we shall examine several rescaled single scalar field
models of inflation regarding their inflationary phenomenology
when compared to the ACT and the updated Planck/BICEP data.

\section{Scalar Field Inflationary Models and the ACT Constraints}

In the following we will examine for which values of the
dimensionless parameter $\alpha$ do some inflationary models
appearing in the Planck 2018 release of inflation
\cite{Planck:2018jri} and in Ref. \cite{Martin:2013tda}, comply
with the recent ACT (\ref{act}) and updated Planck/BICEP data of
Eq. (\ref{planck}).

\subsection{Rescaled D-Brane ($p=2$) Inflation}

\begin{figure}[t]
    \centering
    \includegraphics[width=0.45\textwidth]{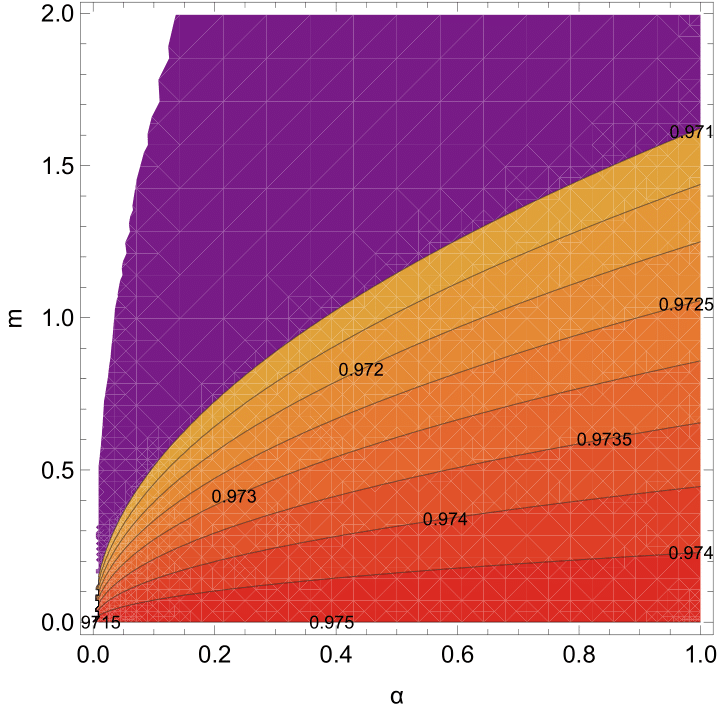}
    \hfill
    \includegraphics[width=0.45\textwidth]{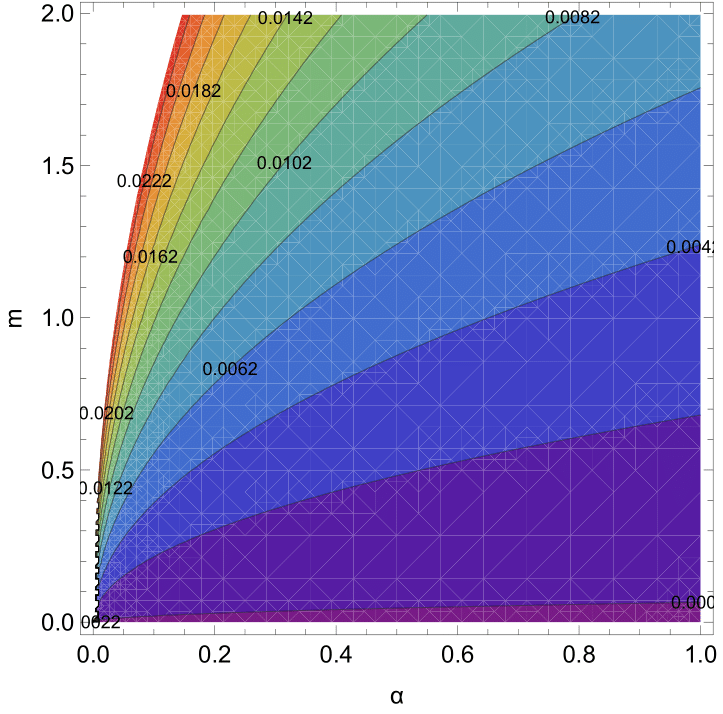}
    \caption{Contour plot for the spectral index of primordial scalar curvature
perturbations $n_s$ (left plot) and the tensor-to-scalar ratio $r$
(right plot) for $\alpha = [0, 1]$, $m = [10^{-6}, 10^{0.3}]$ and
$N = 60$ for the D-Brane Model ($p=2$).}
    \label{fig:db260}
\end{figure}

\begin{figure}[t]
    \centering
    \includegraphics[width=0.45\textwidth]{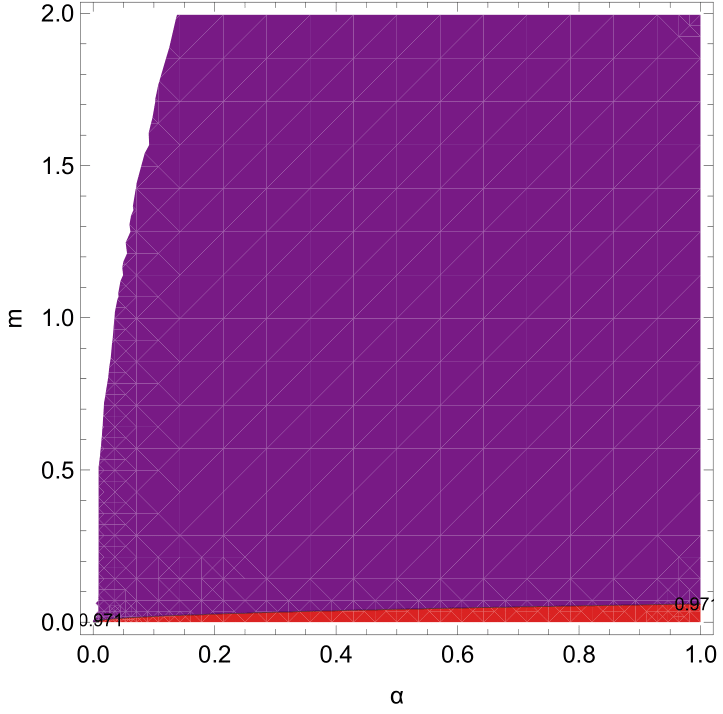}
    \hfill
    \includegraphics[width=0.45\textwidth]{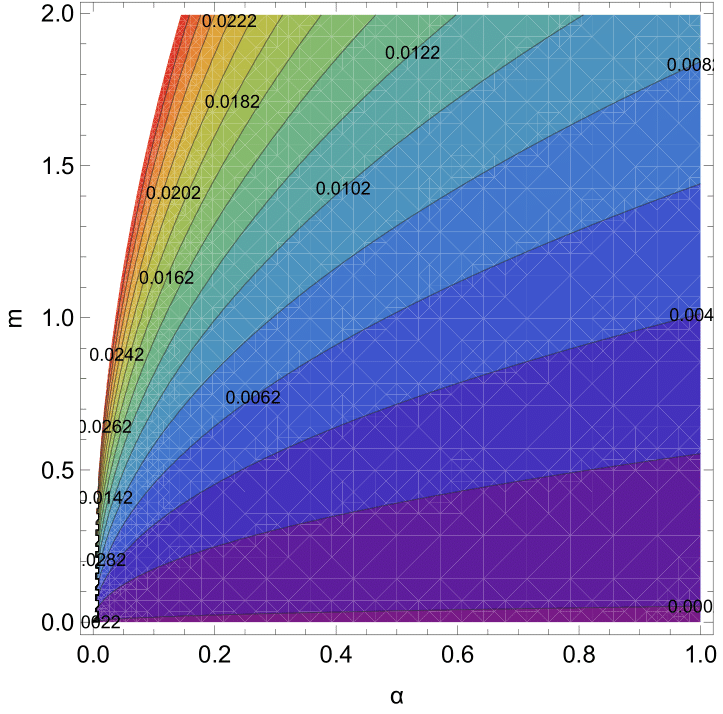}
    \caption{Contour plot for the spectral index of primordial scalar curvature
perturbations $n_s$ (left plot) and the tensor-to-scalar ratio $r$
(right plot) for $\alpha = [0, 1]$, $m = [10^{-6}, 10^{0.3}]$ and
$N = 52$ for the D-Brane Model ($p=2$). Here we are at the
borderline of the constraint for $n_s$.}
    \label{fig:db252}
\end{figure}

\begin{figure}
\centering
\includegraphics[width=28pc]{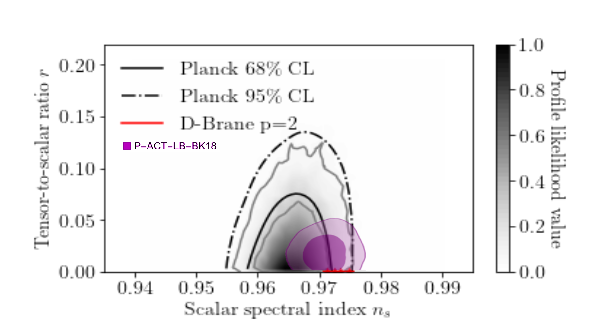}
\caption{Marginalized curves of the Planck 2018 data and the
rescaled D-Brane gravity model $p=2$, confronted with the ACT
data, the Planck 2018 data, and the updated Planck/BICEP
constraints on the tensor-to-scalar ratio. We choose $\alpha=0.8$
and $m=0.1$ and $N$ in the range $N=[50,60]$.
}\label{actplotmodel1}
\end{figure}

The first model we are examining is a D-Brane model with an
exponent $p=2$, \cite{Planck:2018jri},
\begin{equation}\label{D-Brane2}
V(\phi)=\Lambda ^4 \left(1-\left(\frac{m}{\kappa\phi}\right)^2\right),
\end{equation}
where $\Lambda$ has dimensions of mass [m] and $m$ is a
dimensionless parameter taking values in $[10^{-6}, 10^{0.3}]$.
Using equation (\ref{eps1 to eps}) we can obtain the first
slow-roll index for this model,
\begin{equation}\label{Db2e1}
    \epsilon_1 \simeq \frac{2 \alpha  m ^4}{\kappa^6 \phi ^6}.
    \end{equation}
By solving the equation $\epsilon_1(\phi_f)=1$ we find the value
of $\phi$ at the end of the inflationary era,
\begin{equation}\label{Db2ff}
    \phi_f=\frac{\sqrt[6]{2} \sqrt[6]{\alpha } m ^{2/3}}{\kappa},
\end{equation}
and using the integral in (\ref{N}) we get the value of $\phi$ at
the beginning of this era as well to be,
\begin{equation}\label{D2fi}
    \phi_i=\frac{\sqrt[4]{2^{2/3} \alpha ^{2/3} m ^{8/3}+8 \alpha  m ^2 N}}{\kappa}.
\end{equation}
Now, the expressions of the spectral index and the
tensor-to-scalar ratio at this point at leading order in $1/N$
are,
\begin{equation}\label{Db2ns}
    n_s \simeq 1-\frac{3}{2 N}-\frac{3 m}{2 \sqrt{2 \alpha } N^{3/2}}+\frac{3 m ^{2/3}}{8 \sqrt[3]{2} \sqrt[3]{\alpha } N^2},
\end{equation}
\begin{equation}\label{Db2r}
   r \simeq \frac{\sqrt{2} m }{\sqrt{\alpha } N^{3/2}}+\frac{m^2}{\alpha N^2}.
\end{equation}\\
Considering that both $n_s$ and $r$ should comply with the ACT
constraints in (\ref{ACT}) we should search for the values of
$\alpha$ and $m$ that satisfy these. We assume that we are at the
end of the inflationary era, so the number of e foldings is $N \in
[50, 60]$. Given that, we construct the plots of
Figs.~\ref{fig:db260},~\ref{fig:db252}.
\\
These plots have been restricted to show only the regions that
$n_s=0.9743 \pm 0.0034$ and $r<0.034$. In order to examine the
viability of this D-Brane potential, we tested a singular value
for $\alpha$ and m ($\alpha=0.8$ and $m=0.1$), for $N=53, 57, 60$.
The pairs of $n_s$ and $r$ are shown below:
\begin{table}[hbt!]
    \centering
    \def\arraystretch{1.5}
       \caption{Values of $N$, $a$, $m$, $n_s$, $r$}
      \scriptsize
    \begin{tabular}{c|c|c|c|c}
        \hline
        e-foldings&$a$&$m$&$n_s$&$r$\\
        \hline\hline
        53&0.8&0.1&0.971413&0.000414235\\
        \hline
        57&0.8&0.1&0.973428&0.000371263\\
        \hline
        60&0.8&0.1&0.974762&0.000343679\\
        \hline
    \end{tabular}
    \label{tab:placeholder}
\end{table}
Generally, $\alpha=[0, 1]$ so that the rescaling leads effectively
to a strong Einstein-Hilbert gravity. In order to have a concrete
idea of the viability of the model, in Fig. \ref{actplotmodel1} we
confront directly the model at hand with the ACT data, the Planck
2018 data and the updated Planck/BICEP data, choosing $\alpha=0.8$
and $m=0.1$, and also $N$, the $e$-foldings number, in the range
$N=[50,60]$. As it can be seen, the model is well compatible with
both the ACT and the updated Planck/BICEP data.

\subsection{Rescaled D-Brane Inflation ($p=4$)}

\begin{figure}[t]
    \centering
    \includegraphics[width=0.45\textwidth]{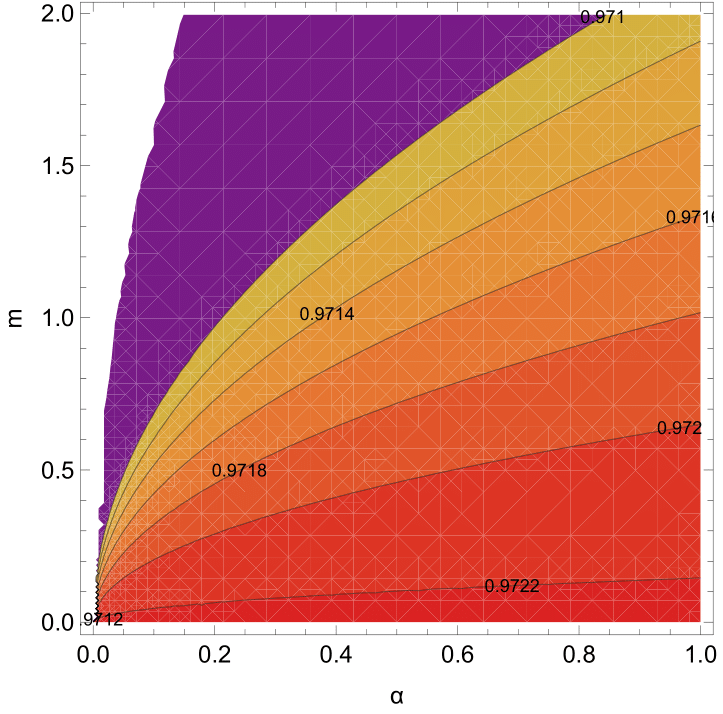}
    \hfill
    \includegraphics[width=0.45\textwidth]{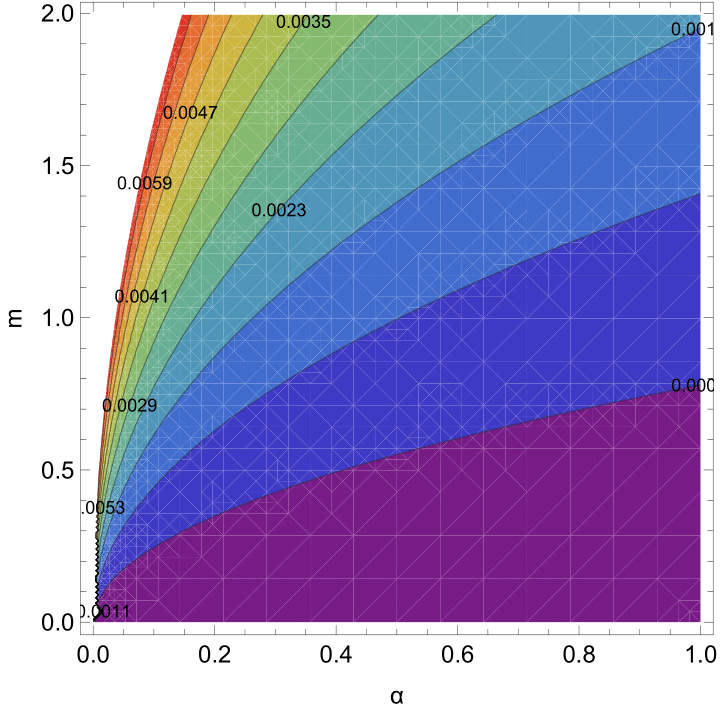}
    \caption{Contour plot for the spectral index of primordial scalar curvature
perturbations $n_s$ (left plot) and the tensor-to-scalar ratio $r$
(right plot) for $\alpha = [0, 1]$, $m = [10^{-6}, 10^{0.3}]$ and
$N = 60$ for the D-Brane Model (p=4).}
    \label{fig:db460}
\end{figure}

\begin{figure}[t]
    \centering
    \includegraphics[width=0.45\textwidth]{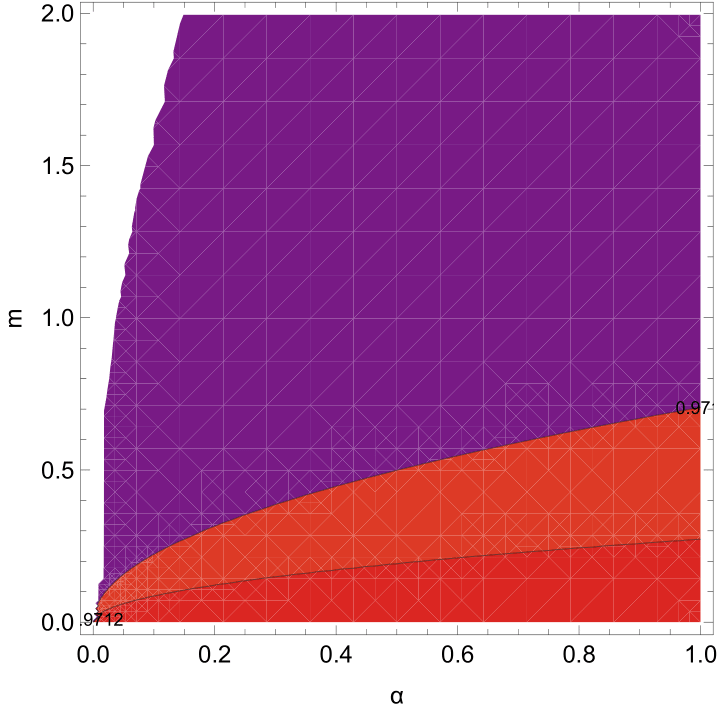}
    \hfill
    \includegraphics[width=0.45\textwidth]{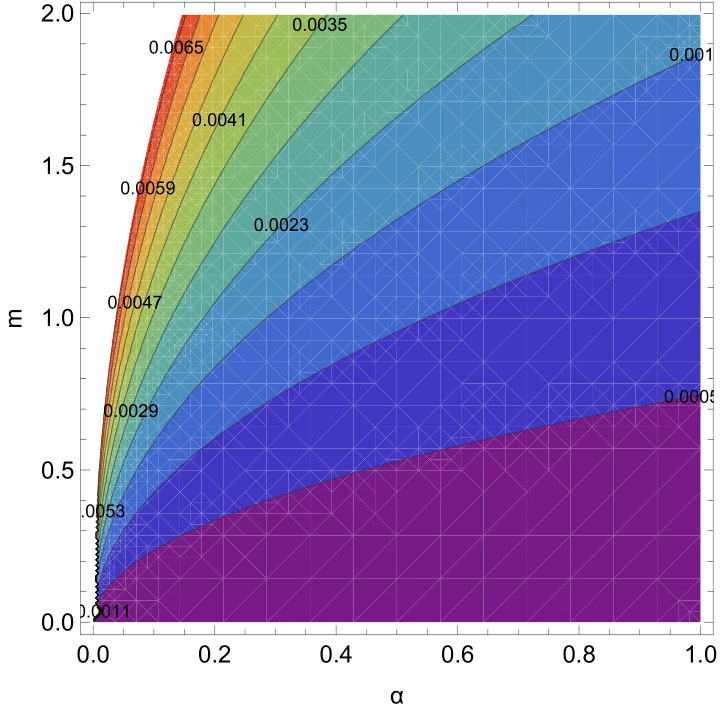}
    \caption{Contour plot for the spectral index of primordial scalar curvature
perturbations $n_s$ (left plot) and the tensor-to-scalar ratio $r$
(right plot) for $\alpha = [0, 1]$, $m = [10^{-6}, 10^{0.3}]$ and
$N = 58$ for the D-Brane Model (p=4). Here we are at the
borderline of the constraint for $n_s$.}
    \label{fig:db458}
\end{figure}

We proceed to examine another D-Brane inflation model with an
exponent $p=4$, \cite{Planck:2018jri},
\begin{equation}\label{DBrane4}
V(\phi)=\Lambda ^4 \left(1-\left(\frac{m}{\kappa\phi}\right)^4\right),
\end{equation}
where $\Lambda$ has dimensions of mass $[m]$ and $m$ is a
dimensionless parameter. For this model the first slow-roll index
is,
\begin{equation}\label{Dbrane4e1}
    \epsilon_1 \simeq \frac{8 \alpha  m ^8}{\kappa^{10} \phi
    ^{10}}.
    \end{equation}
We can obtain the value of $\phi$ at the end of inflation by
solving the equation $\epsilon_1 (\phi_f)=1$ and, thus, we get
that,
\begin{equation}\label{Dbrane4ff}
    \phi_{f}=\frac{2^{3/10} \sqrt[10]{\alpha } m ^{4/5}}{\kappa}.
\end{equation}
From the integral of (\ref{N}) solved with respect to $\phi_i$ we find,
\begin{equation}\label{Dbrane4fi}
    \phi_i=\frac{\sqrt[6]{2} \sqrt[6]{2^{4/5} \alpha ^{3/5} m ^{24/5}+12 \alpha  m ^4 N}}{\kappa},
\end{equation}
and we may obtain the expressions of the spectral index $n_s$ and
the tensor-to-scalar ratio $r$ at the beginning of the
inflationary era at leading order in $1/N$,
\begin{equation}\label{DBranens}
    n_s \simeq 1-\frac{5}{3 N}-\frac{11 m ^{4/3}}{12\ 3^{2/3} N^{5/3} a^{2/3}}+\frac{5 m ^{4/5}}{18 \sqrt[5]{2} \alpha ^{2/5} N^2},
\end{equation}
\begin{equation}\label{Dbraner}
   r \simeq \frac{4   m ^{4/3} }{3\ 3^{2/3} \alpha^{2/3}  N^{5/3}}.
\end{equation}
As shown in the plots Figs.~\ref{fig:db460},~\ref{fig:db458} that
depict the behavior of $n_s$ and $r$ for $N \simeq 60$ and $N
\simeq 58$, respectively, there is some variety to the values for
$\alpha$ and $m$ that the constraints for $n_s$ and $r$ in
(\ref{ACT}) are met for $N\in [58, 60]$. As an example, for
further clarification, we select one value for $\alpha$ and $m$,
specifically $\alpha=0.001$ and $m=0.065$. In result we obtain
$n_s=0.971057$ and $r=0.00182176$, which are in accordance with
the most recent ACT and updated BICEP/Planck data.
\\

\begin{figure}
\centering
\includegraphics[width=28pc]{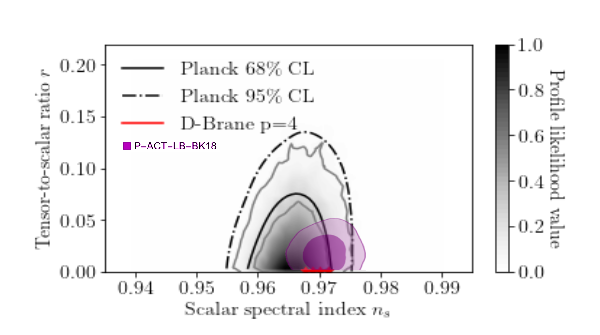}
\caption{Marginalized curves of the Planck 2018 data and the
rescaled D-Brane gravity model ($p=4$), confronted with the ACT
data, the Planck 2018 data, and the updated Planck/BICEP
constraints on the tensor-to-scalar ratio. We choose $\alpha=0.5$
and $m=0.9$ and $N$ in the range $N=[50,60]$.
}\label{actplotmodel2}
\end{figure}

In order to have a concrete idea of the viability of the model at
hand, in Fig. \ref{actplotmodel2} we confront directly the model
at hand with the ACT data, the Planck 2018 data and the updated
Planck/BICEP data, choosing $\alpha=0.5$ and $m=0.9$, and also
$N$, the $e$-foldings number, in the range $N=[50,60]$. As it can
be seen, the model is marginally compatible with the ACT, but
fully compatible with the updated Planck/BICEP data.

\subsection{Rescaled Einstein-Frame Plateau Potentials}

\begin{figure}[t]
    \centering
    \includegraphics[width=0.45\textwidth]{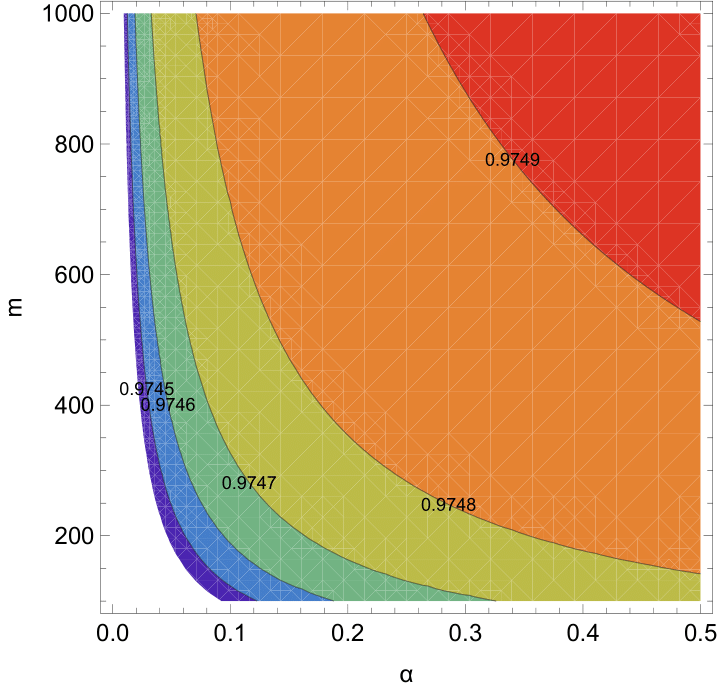}
    \hfill
    \includegraphics[width=0.45\textwidth]{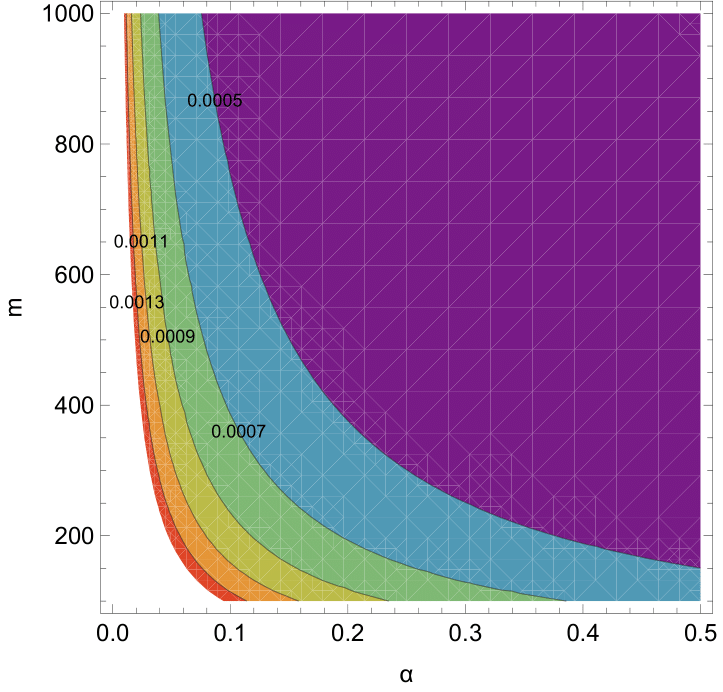}
    \caption{Contour plot for the spectral index of primordial scalar curvature
perturbations $n_s$ (left plot) and the tensor-to-scalar ratio $r$
(right plot) for $\alpha = [0, 1]$, $\xi = [10^{2}, 10^{4}]$ and
$N = 60$ for the Einstein Frame Plateau potential.}
    \label{fig:efp60}
\end{figure}

\begin{figure}[t]
    \centering
    \includegraphics[width=0.45\textwidth]{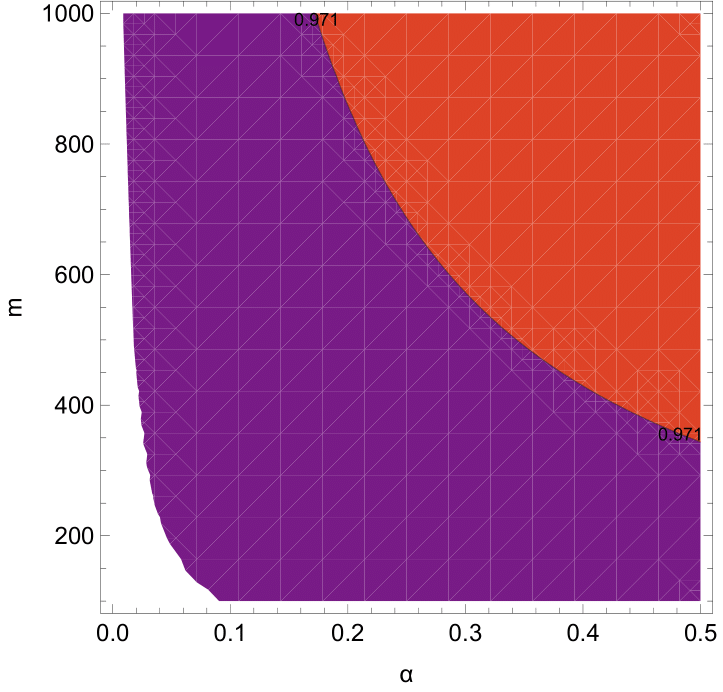}
    \hfill
    \includegraphics[width=0.45\textwidth]{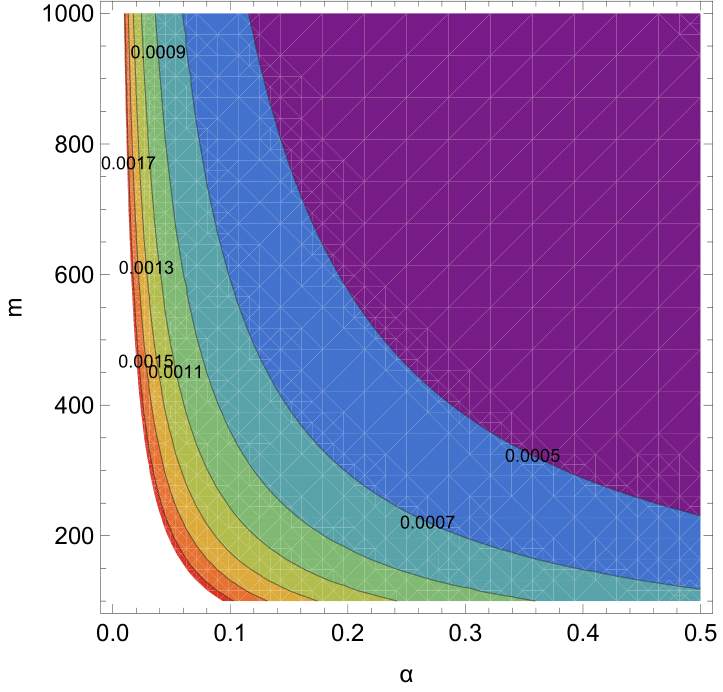}
    \caption{Contour plot for the spectral index of primordial scalar curvature
perturbations $n_s$ (left plot) and the tensor-to-scalar ratio $r$
(right plot) for $\alpha = [0, 1]$, $\xi = [10^ {2}, 10^{4}]$ and
$N = 52$ for the Einstein Frame Plateau potential. Here we are at
the borderline for the constraint of $n_s$.}
    \label{fig:efp52}
\end{figure}

This inflationary model belongs to the family of Einstein-frame
plateau potentials generated by a pole in the non-minimal coupling
arising when the Jordan-frame conformal factor has a second-order
pole. It is described by the following potential,
\begin{equation} \label{Einstein_potential}
V(\phi)=\Lambda^4{\left(1-{\left(\frac{1}{\xi(\kappa\phi)^2}\right)}\right)}^2 ,
\end{equation}
where $\Lambda$ has dimensions of mass $[m]$ and $\xi$ is a
dimensionless constant that generally takes values in $[10^{2} ,
10^{4}]$. For this potential, the first slow-roll index takes
the form,
\begin{equation}\label{Nate1}
    \epsilon_1=\frac{8\alpha}{\kappa^2\phi^2(-1+\xi\kappa^2\phi^2)^2} .
\end{equation}
We obtain $\phi$ at the end of the inflationary era from the
equation $\epsilon_1(\phi_f)=1$ and we get,
\begin{equation}\label{Natff}
    \phi_f=\frac{\sqrt{2}\alpha^{1/6}}{\xi^{1/3}\kappa  }.
\end{equation}
Proceeding, we use the integral that gives the number of
$e$-foldings $N$ from (\ref{N}) to find $\phi_i$, which is,
\begin{equation}\label{Natfi}
  \phi_i=\frac{\sqrt{2}{\left(\frac{\alpha^{2/3}}{\xi^{1/3}}+4\alpha N\right)}^{1/4}}{\xi^{1/4}\kappa  }.
\end{equation}
Our interest lies in the spectral index $n_s$ and the
tensor-to-scalar ratio $r$ so, for this model at $\phi_i$ their
expressions are,
\begin{equation}\label{Einns}
    n_s=1-\frac{3}{2N}-\frac{7}{8\sqrt{\alpha\xi}N^{3/2}}-\frac{-11+12\alpha^{2/3}\xi^{2/3}}{32\alpha\xi N^2},
\end{equation}
\begin{equation}\label{Einr}
    r=\frac{2}{\sqrt{\alpha\xi}N^{3/2}}
\end{equation}
We seek for the values of $\alpha$ and $\xi$ that allow $n_s$ and
$r$ to lie within the range of the ACT constraints in (\ref{ACT}).
The plots in Figs.~\ref{fig:efp60},~\ref{fig:efp52} present the
behavior of $n_s$ and $r$ for $N \simeq 60$, $N \simeq 57$, $N
\simeq 53$ and $N \simeq 52$, respectively, while lying in the
constraints region. Let us provide an example, for this potential,
for $\alpha=0.1$ and $\xi=130$, in which we obtain the following
table of pairs of $n_s$ and $r$:
\begin{table}[hbt!]
    \centering
    \def\arraystretch{1.5}
       \caption{Values of $N$, $a$, $\xi$, $n_s$, $r$}
      \scriptsize
    \begin{tabular}{c|c|c|c|c}
        \hline
        e-foldings&$a$&$\xi$&$n_s$&$r$\\
        \hline\hline
        53&0.1&130&0.971117&0.00146501\\
        \hline
        57&0.1&130&0.973161&0.00131266\\
        \hline
        60&0.1&130&0.974515 &0.00121489\\
        \hline
    \end{tabular}
    \label{tab:placeholder}
\end{table}\\
for the end of the inflationary era.
\\

\begin{figure}
\centering
\includegraphics[width=28pc]{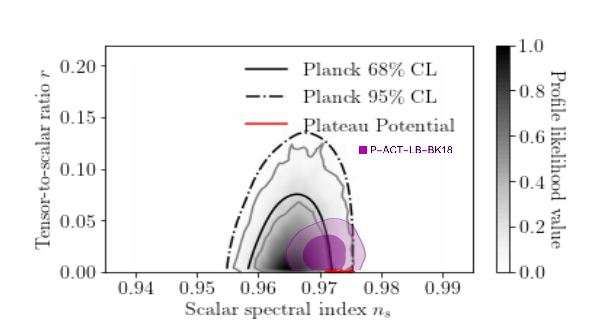}
\caption{Marginalized curves of the Planck 2018 data and the
rescaled Einstein frame plateau gravity model, confronted with the
ACT data, the Planck 2018 data, and the updated Planck/BICEP
constraints on the tensor-to-scalar ratio. We choose $\alpha=0.7$
and $m=150$ and $N$ in the range $N=[50,60]$.
}\label{actplotmodel3}
\end{figure}

In order to have a concrete idea of the viability of the model at
hand, in Fig. \ref{actplotmodel3} we confront directly the model
at hand with the ACT data, the Planck 2018 data and the updated
Planck/BICEP data, choosing $\alpha=0.7$ and $m=150$, and also
$N$, the $e$-foldings number, in the range $N=[50,60]$. As it can
be seen, the model is well compatible with both the ACT and the
updated Planck/BICEP data.

\subsection{Rescaled Exponential T-model ($n=1$)}

\begin{figure}[t]
    \centering
    \includegraphics[width=0.45\textwidth]{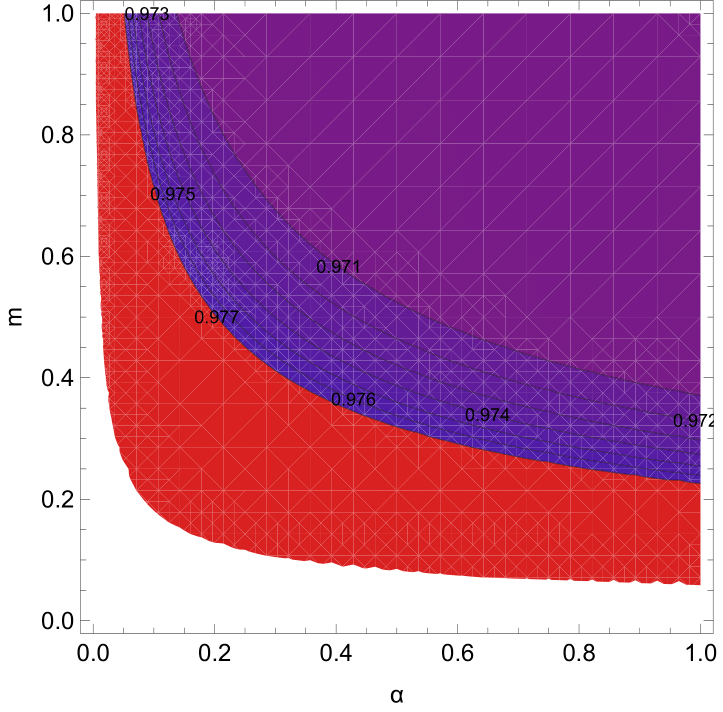}
    \hfill
    \includegraphics[width=0.45\textwidth]{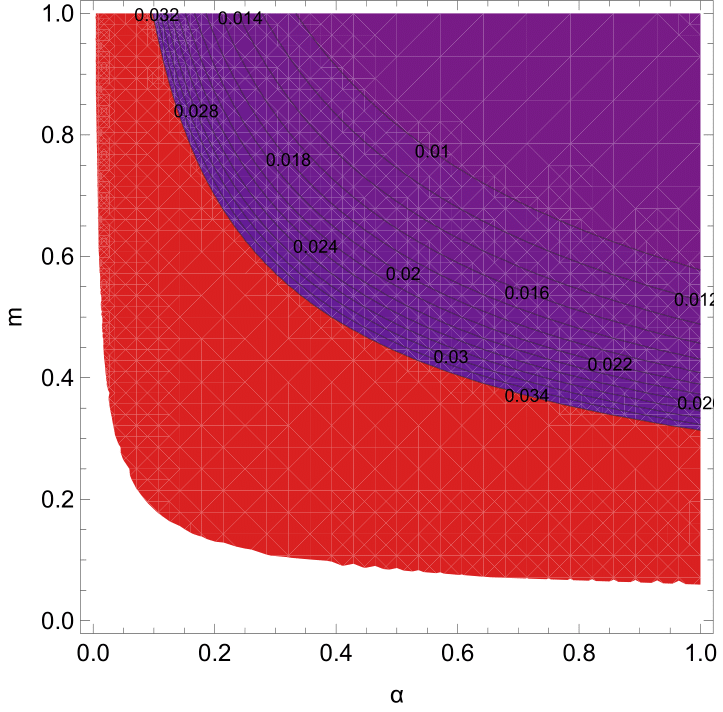}
    \caption{Contour plot for the spectral index of primordial scalar curvature
perturbations $n_s$ (left plot) and the tensor-to-scalar ratio $r$
(right plot) for $\alpha = [0, 1]$, $m = [10^{-2}, 10]$ and $N =
60$ for the exponential T-Model ($n=2$).}
    \label{fig:etm60}
\end{figure}

\begin{figure}[t]
    \centering
    \includegraphics[width=0.45\textwidth]{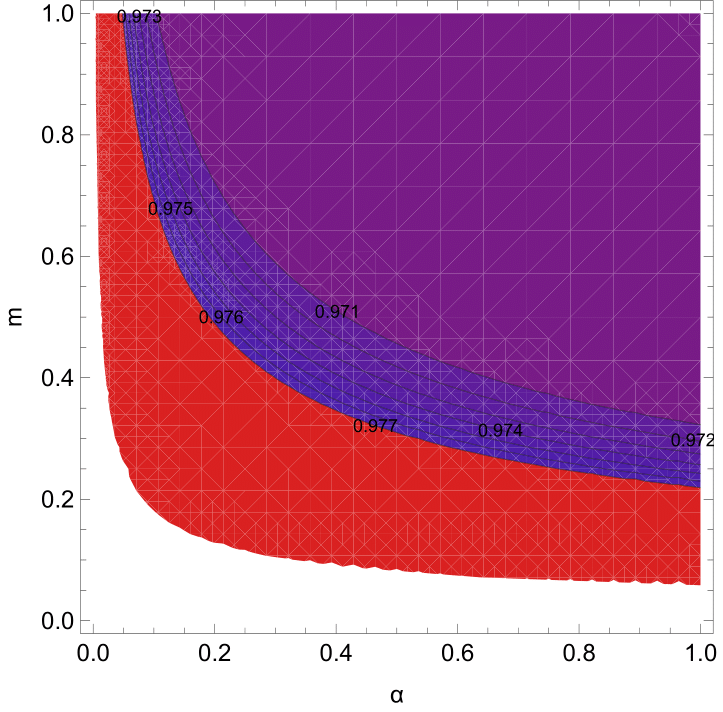}
    \hfill
    \includegraphics[width=0.45\textwidth]{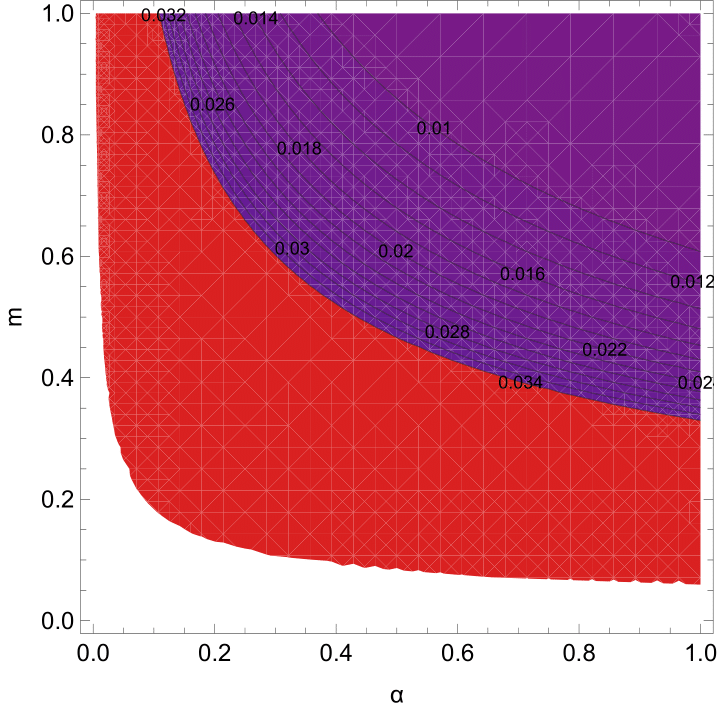}
    \caption{Contour plot for the spectral index of primordial scalar curvature
perturbations $n_s$ (left plot) and the tensor-to-scalar ratio $r$
(right plot) for $\alpha = [0, 1]$, $m = [10^{-2}, 10]$ and $N =
57$ for the exponential T-Model ($n=2$).Here we are at the
borderline for the constraint of $n_s$.}
    \label{fig:etm57}
\end{figure}

The next inflationary model to study belongs to the family of
T-model $\alpha$-attractors (exponential plateau inflation), with
$n=1$,
\begin{equation}\label{ETM}
 V(\phi)=\Lambda ^4 {\left(1+e^{-m\sqrt{2/3}\kappa\phi}\right)^{-2}},
\end{equation}
where, as usual, $\Lambda$ has dimensions of mass $[m]$ and $m$ is
a dimensionless parameter with values $m\in [10^{-2}, 10]$. This
exponential form corresponds to a large field inflation, from an
approximation of the original T-Model:
\begin{equation}
   V(\phi)=\Lambda ^4 \tanh ^{2n}{\left(\frac{m\kappa\phi}{\sqrt{6}}\right)}.
\end{equation}
The first slow-roll index in this case is,
\begin{equation}
\label{HPe1}
    \epsilon_1 = \frac{4 \alpha m^2 }{3 \left(e^{\sqrt{2/3}m\kappa\phi}\right)^2}.
\end{equation}
Once more, we solve the equation $\epsilon_1(\phi_f)=1$ and we obtain $\phi$ at the end of inflation,
\begin{equation}\label{HPff}
    \phi_f=\frac{\sqrt{\frac{3}{2}} Log\left(\frac{4\alpha m^2}{3}\right)}{2l\kappa},
\end{equation}
and from the integral of (\ref{N}) we find the initial $\phi_i$,
\begin{equation}\label{HPfi}
    \phi_i=\frac{\sqrt{\frac{3}{2}} Log\left(\frac{2}{3} \left(\sqrt{3} \sqrt{\alpha m^2}+2\alpha l^2 N\right)\right)}{m\kappa},
\end{equation}
where $N$ is the number of $e$-foldings the Universe has grown
during the inflationary era. The spectral index of the primordial
curvature perturbations $n_s$ and the tensor-to-scalar ratio $r$
for this potential are,
\begin{align}\label{HPns}
    n_s=1-\frac{2}{N}+\frac{3+2\sqrt{3}\sqrt{\alpha m^2}}{2\alpha m^2 N^2},
\end{align}
\begin{equation}\label{HPr}
   r=\frac{12}{\alpha m^2 N^2}.
\end{equation}
This model can provide us with viable inflation for a variety of
$\alpha$ $(\alpha \in (0,1))$ and $m$ values. Furthermore, $n_s$
and $r$ comply with the ACT constraints when $N \in [57,60]$, as
it can be seen in the plots in
Figs.~\ref{fig:etm60},~\ref{fig:etm57}
\\

\begin{figure}
\centering
\includegraphics[width=28pc]{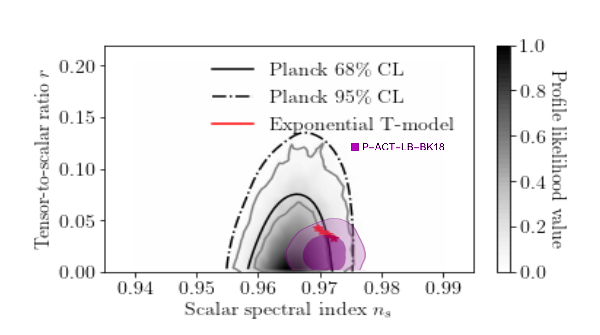}
\caption{Marginalized curves of the Planck 2018 data and the
rescaled exponential T-Model ($n=2$) gravity model, confronted
with the ACT data, the Planck 2018 data, and the updated
Planck/BICEP constraints on the tensor-to-scalar ratio. We choose
$\alpha=0.1$ and $m=1$ and $N$ in the range $N=[50,60]$.
}\label{actplotmodel4}
\end{figure}

Of course, we can display an example for $\alpha=0.1$ and $m=1$
and obtain $n_s=0.972355$ and $r = 0.033$, which fit in the ACT
observational constraints. In order to have a concrete idea of the
viability of the model at hand, in Fig. \ref{actplotmodel4} we
confront directly the model at hand with the ACT data, the Planck
2018 data and the updated Planck/BICEP data, choosing $\alpha=0.1$
and $m=0.1$, and also $N$, the $e$-foldings number, in the range
$N=[50,60]$. As it can be seen, the model is very marginally
compatible with both the ACT and the updated Planck/BICEP data.

\section{Conclusions}

In this article we considered an effective theory of gravity in
the context of single scalar canonical field theory, in which
primordially gravity was stronger than Einstein-Hilbert gravity.
This theory is realized by terms of the form $\sim \alpha R$, with
$\alpha$ in the range $(0,1)$ and these are generated effectively
by an $F(R)$ theory of gravity in the large curvature regime which
affects the inflationary era. We considered the viability of the
effective single scalar field theory when compared to the latest
ACT constraint on the spectral index of the scalar perturbations
and the updated Planck/BICEP constraint on the tensor-to-scalar
ratio. For our analysis we chose four popular single scalar field
theory models, namely D-Brane inflation with exponent $p=2$,
D-Brane inflation with exponent $p=4$, exponential T-model
inflation and an Einstein frame pole inflation model. The D-Brane
inflation with exponent $p=2$ model and the Einstein frame pole
inflation model are fully compatible with the ACT data and the
updated Planck/BICEP constraint, but the exponential T-model and
the D-Brane inflation with exponent $p=4$ model were found
marginally compatible with the ACT constraint or the updated
Planck/BICEP constraint. Thus the rescaled gravity framework
offers the possibility that models of inflation which are excluded
by the ACT constraints, may become compatible with it if the
Einstein-Hilbert gravity was stronger primordially during
inflation. We believe that this possibility may offer new insights
for inflationary phenomenology.

\end{document}